\documentclass[11pt,a4paper]{article}

\usepackage{amsmath,amssymb}
\usepackage{epsfig,graphicx}
\usepackage{subfigure}
\usepackage{graphicx}
\usepackage{rotating}
\usepackage{cancel}
\usepackage{bm}
\usepackage{color}
\usepackage{comment}
\usepackage{cite}
\usepackage{psfrag}
\usepackage{appendix}

\newcommand{\mg}{m}

\newcommand{\degree}{\ensuremath{^\circ}}

\renewcommand\({\left(}
\renewcommand\){\right)}
\renewcommand\[{\left[}

\newcommand{\exclude}[1]{}

\def\beq{\begin{equation}}
\def\eeq{\end{equation}}

\begin{document}
\numberwithin{equation}{section}
\title{{\normalsize  \mbox{}\hfill MPP-2013-216}\\
\vspace{2.5cm} 
\Large{\textbf{An  Antenna for Directional Detection of WISPy Dark Matter
\vspace{0.5cm}}}}

\author{Joerg Jaeckel$^{1}$, and Javier Redondo$^{2,3}$\\[2ex]
\small{\em $^1$Institut f\"ur theoretische Physik, Universit\"at Heidelberg, Philosophenweg 16, 69120 Heidelberg, Germany}\\[0.5ex]
\small{\em $^2$Arnold Sommerfeld Center, Ludwig-Maximilians-Universit\"at, Munich, Germany}\\[0.5ex]
\small{\em $^3$Max-Planck-Institut f\"ur Physik, Munich, Germany}\\[0.5ex]
}

\date{}
\maketitle

\begin{abstract}
It is an intriguing possibility that the cold dark matter of the Universe may consist of very light and very weakly interacting particles such as axion(-like particles) and hidden photons. This opens up (but also requires) new techniques for direct detection. One possibility is to use reflecting surfaces to facilitate the conversion of dark matter into photons, which can be concentrated in a detector with a suitable geometry. In this note we show that this technique also allows for directional detection and inference of the full vectorial velocity spectrum of the dark matter particles. We also note that the non-vanishing velocity of dark matter particles is relevant for the conception of (non-directional) discovery experiments and outline relevant features.
\noindent
\end{abstract}

\newpage

\section{Introduction}
Dark matter is still one of the most important mysteries of physics. These new particles contribute about 27\% of the total energy content of the Universe~\cite{Ade:2013zuv} and still their nature is unknown~\cite{Bertone:2010zz}. Clarification requires input from experiment, preferably by the direct detection of dark matter particles.

Two leading candidates are the QCD axion~\cite{Peccei:1977hh,Weinberg:1977ma,Wilczek:1977pj,Preskill:1982cy,Abbott:1982af,Dine:1982ah,Sikivie:1982qv,Hiramatsu:2010yn} (see~\cite{Kawasaki:2013ae} for a recent review), and weakly interacting massive particles (WIMPs)~\cite{Bertone:2004pz}.
The former are motivated by the solution to the strong CP problem, the latter often arise in extensions of the Standard Model such as Supersymmetry~\cite{Jungman:1995df}. 

These two candidates are representative of two quite different scenarios. Axion dark matter consists of very light ($\lesssim\,{\rm eV}$) and very weakly particles produced non-thermally, e.g. by the misalignment mechanism. 
Indeed as has recently been argued that a whole class of very weakly interacting slim particles (WISPs, see~\cite{Jaeckel:2010ni} for a review) may comprise the dark matter, originating from such a mechanism~\cite{Piazza:2010ye,Nelson:2011sf,Arias:2012az}. Beyond the axion this class contains more general axion-like particles as well as extra U(1) gauge bosons, often called hidden photons, both well motivated from extensions of the Standard Model based on field and string theory, see~\cite{Svrcek:2006yi
} and \cite{Dienes:1996zr}. The aim of this work is to suggest a highly sensitive way for the directional detection of this type of dark matter.

WIMPs on the other hand are typically quite heavy ($\gtrsim 1\,{\rm GeV}$) and produced from collisions in the thermal bath.
Due to their different properties it is clear that WIMPs and WISPs require entirely different search strategies.
WIMPs are usually sought after by looking for WIMP-nucleon collisions~\cite{Baudis:2012ig}, measuring the energy of the nucleon recoiling against the WIMP. The recoil energy grows with the mass of the WIMP colliding with the nucleus and current detector thresholds therefore require that the particles have masses $\gtrsim 1\,{\rm GeV}$ much higher than those of WISPs.
For the detection of WISPs one can, however, use that fact that axions and axion-like particles can be converted into photons in the presence of magnetic fields~\cite{Sikivie:1983ip}. For hidden photons this happens even in vacuum~\cite{Arias:2012az}. These photons can then be detected. 

One setup for the detection of WISPy DM is the axion-haloscope~\cite{Sikivie:1983ip} as currently employed in the ADMX experiment~\cite{Asztalos:2009yp} (with several others in planning). Basically it consists of a cavity which amplifies the WISP--photon conversion when the mass of the  DM particle is equal (within the bandwidth) to the resonance frequency of the cavity, coupled to a highly sensitive photon detector. This technique is currently most advanced in the range of microwave frequencies corresponding to masses of the order of $\sim(1-100)\,\mu{\rm eV}$. Although this technique is extremely sensitive it suffers from the fact that resonant enhancement only works when the mass of the DM particle and the frequency of the cavity coincide. Covering a mass range therefore requires a time-consuming scan through the mass, by tuning the resonance frequency of the cavity.

Recently we have suggested a non-resonant broadband search strategy for WISPy DM~\cite{Horns:2012jf} using suitably formed reflective surfaces functioning as a``dish antenna'' for WISPy DM. This technique employs that reflective surfaces can convert WISPs into photons. Given suitable geometry, a spherical cap, these photons are concentrated in the centre where they can be detected. The advantages of this technique are that the same reflector can be used in a broad range of masses (no need to tune anything) at the same time by using a suitable broadband detector and that the setup is scalable by going to larger antennas.

Once a DM signal has been established in a direct detection experiment one would like to obtain additional information on the kinetic energy and eventually the full 3-dimensional velocity spectrum of the dark matter particles. This could give us vital information on the formation of our galaxy, as well as on structure formation in general. 
The most sensitive WIMP detectors currently offer little or no directional sensitivity. While techniques are being developed, directional sensitivity will require a whole new set of detectors and is likely to happen many years after a signal has first been found. 

For WISPs the prospects for directional detection are brighter. Once a signal has been established in a cavity experiment, the mass of the particle is known with high precision. Measuring at this frequency (without the need to change the apparatus) for just a little bit longer microwave detectors can give a high resolution spectrum of the kinetic energy. Beyond this it has been proposed that one could gain directional information by employing a suitably shaped cavity~\cite{Irastorza:2012jq}.

The purpose of this note is to show that for a dish antenna search for WISPy DM the full 3-D velocity distribution can be inferred, with little or no change to the experiment. 

\section{Directional detection with a dish antenna}\label{directsect}
In this section we will discuss how directional detection can be achieved using a dish antenna. We will concentrate on the simplest case of hidden photons, but the situation for axion-like particles is completely analogous as already discussed in~\cite{Horns:2012jf}.

\subsection{Hidden photon dark matter}\label{HPDM}
To start let us recall the basic equations. In a suitable field basis (cf.~\cite{Horns:2012jf}) the Lagrangian describing hidden photons (HPs henceforth) is given by,
\begin{equation}
\label{HPlagrangian}
\mathcal L= -\frac{1}4F_{\mu\nu}F^{\mu\nu}-\frac{1}4 X_{\mu\nu} X^{\mu\nu}+\frac{m^2}2 (X_\mu X^\mu-2\chi A_{\mu}X^{\mu}+\chi^2 A_{\mu}A^{\mu})+ J^\mu A_\mu , 
\end{equation}
where $A_\mu,X_\mu$ are the photon and HP fields with field strengths $F_{\mu\nu},X_{\mu\nu} $, $m$ is the HP mass and $\chi$ is a tiny kinetic mixing parameter. 
In the low energy limit, HPs do not couple directly to any particle of the standard model, their sole interaction is through the small mixing $\chi$ with photons. 
The equations of motion for plane waves with frequency $\omega$ and wavenumber $p$ are, 
\begin{equation}
\left[(\omega^2-p^2)\left(
\begin{array}{cc}
1  & 0     \\
0 &  1   \\  
\end{array}
\right)
-
\mg^2\left(
\begin{array}{cc}
\chi^2 & -\chi     \\
 -\chi &  1    \\  
\end{array}
\right)
\right]
\left(
\begin{array}{c}
A^{\mu}\\
X^{\mu}
\end{array}
\right)=
\left(\begin{array}{c}
0\\
0
\end{array}
\right).
\end{equation}
For homogeneous solutions, i.e. $p=0$ we can 
achieve $X^{0}=A^{0}=0$ by a suitable a suitable gauge choice. For $p\neq 0$ the Lorentz condition for the massive gauge field does not always allow to do this exactly. But, as we show in the Appendix the corrections to the final result are of order $v^2=p^2/m^2$ and very small $\sim 10^{-6}$ for the velocities we are interested in. In the following we will therefore use $X^{0}\approx A^{0}\approx0$ and focus on the three-vector components.

The dark matter solution is the only massive eigenstate, which for $\chi\ll 1$ has mass $\simeq \mg$ and is close to the hidden-photon one. 
For a particle traveling with momentum $\mathbf{p}$ and energy $\omega=\sqrt{\mg^2+|\mathbf{p}|^{2}}$ we have,
\begin{equation}
\label{DMsolution}
\left(
\begin{array}{c}
\mathbf{A}\\
\mathbf{X}
\end{array}
\right)
\bigg|_{\rm{DM}}=
\mathbf{X}_{\rm{DM}}(\mathbf{p})
\left(
\begin{array}{c}
-\chi\\
1
\end{array}\right)\exp(-i(\omega t-\mathbf{p}\cdot\mathbf{x})).
\end{equation}

If the total galactic dark matter density at Earth's position $\rho_{\rm CDM, halo}\sim 0.3$ GeV/cm$^3$ is due to HPs, then the phase-space density is huge. 
Since the typical velocity of DM in the galaxy is $v_{\rm typ}\sim O(10^{-3})$, the occupation number is 
\begin{equation}
f \sim \frac{\rho_{\rm CDM, halo}/\mg}{\mg^3v_{\rm typ}^3/3\pi^3}\simeq 10^{33}\(\frac{\rm eV}{\mg}\)^4 \(\frac{10^{-3}}{v_{\rm typ}}\)^3, 
\end{equation}
and we can treat the HP field as classical. 

The integral over HP modes has to add up to the total DM density
\begin{equation}
\label{dmcond}
\rho_{\rm HP}=\frac{\mg^2}{2}
\int\frac{d^{3}\mathbf{p}}{(2\pi)^3}\langle|{\mathbf{X}}_{\rm{DM}}(\mathbf{p})|^2\rangle=\rho_{\rm CDM, halo}. 
\end{equation}
The average is over orientations of ${\mathbf{X}}_{\rm{DM}}$ and takes into account that HPs can have different distributions for their vector field direction. As an example we will consider the same two possibilities as in~\cite{Arias:2012az,Horns:2012jf},
\begin{itemize}
\item[(i)]{$\mathbf{X}_{\rm{DM}}$ is the same everywhere in space.}
\item[(ii)]{The hidden photons behave like a gas of particles with random directions, i.e. we have a mixture of hidden photons ``pointing'' in random directions.}
\end{itemize}
Accordingly, the average is trivial in case (i) and relevant only in case (ii).
In Sect.~\ref{HP-direction} we will briefly comment on the possibility to also measure the distribution of the vector field direction. 

Eq.~\eqref{dmcond} can be nicely understood in terms of a total dark matter amplitude ${\cal X}$ and a velocity 
distribution $f(\mathbf{p})$,
\begin{equation}
\rho_{\rm HP}=\frac{\mg^2}{2}{\cal X}^{2}
\int\frac{d^{3}\mathbf{p}}{(2\pi)^3}\frac{\langle|{\mathbf{X}}_{\rm{DM}}(\mathbf{p})\rangle|^2}{{\cal X}^{2}}
=\frac{\mg^2}{2}{\cal X}^{2}
\int\frac{d^{3}\mathbf{p}}{(2\pi)^3}f(\mathbf{p}),
\end{equation}
where the amplitude is fixed by
\begin{equation}
\label{dmcond2}
\rho_{\rm HP}=\frac{\mg^2}{2}{\cal X}^2=\rho_{\rm CDM, halo}, 
\end{equation}
and $f(\mathbf{p})=\langle|{\mathbf{X}}_{\rm{DM}}(\mathbf{p})\rangle|^2/{\cal X}^{2}$ is the probability density for a DM particle to have momentum $\mathbf{p}$ with normalization
\begin{equation}
\int \frac{d^{3}\mathbf{p}}{(2\pi)^{3}}f(\mathbf{p})=1.
\end{equation}

Although the field is mostly HP-like and thus sterile, it has a small component of an oscillating ordinary electromagnetic field that will allow for its detection,
\begin{equation}
\mathbf{E_{\rm DM}}(\mathbf{p})=\chi \mg \mathbf{X}_{\rm{DM}}(\mathbf{p}).
\end{equation}

\subsection{Directional dish antenna detection}
As discussed in~\cite{Horns:2012jf} the crucial feature of a reflecting surface is that it sets a boundary condition, requiring the ordinary electric field to vanish on the surface in all directions parallel to the surface,
\begin{equation}
\mathbf{E}_{||}|_{\rm surface}=0.
\end{equation}

\subsection*{Plane mirror}
For simplicity let us first consider a perfect plane mirror at $z=0$. 
In this case the hidden photon field on the surface will be cancelled by emitting a suitable outgoing plane wave which is mostly photon-like,
\begin{equation}
\label{photonlike}
\left(\begin{array}{c}
\mathbf{E}\\
\mathbf{E}_{\rm hid}
\end{array}\right)_{\rm out}
=\mathbf{E}_{\rm DM,||}\exp(-i(\omega t-\mathbf{x}\cdot\mathbf{k}))
\left(
\begin{array}{c}
1    \\ 
\chi
\end{array}
\right).
\end{equation}

On the $z=0$ plane, the dark matter field together with the outgoing wave then have to fulfill 
the boundary condition for the electric field components parallel to the plane,
\begin{eqnarray}
\label{nearplane}
\left(\begin{array}{c}
\mathbf{E}\\
\mathbf{E}_{\rm hid}
\end{array}\right)_{\rm total, ||}
&\!\!=\!\!&
\mathbf{E}_{\rm DM, ||}\,\left[
\left(\begin{array}{c}
1 \\
\chi
\end{array}\right)\exp(-i(\omega t-\mathbf{k}\cdot\mathbf{x}))+
\frac{1}{\chi}\left(\begin{array}{c}
-\chi\\
1
\end{array}
\right)\exp(-i(\omega t-\mathbf{p}\cdot\mathbf{x})
\right]_{\mathbf{x}=(x,y,z=0)}
\\\nonumber
&\!\!=\!\!&
\mathbf{\mathbf{E}_{\rm DM, ||}}\frac{1}{\chi}
\left(\begin{array}{c}
0\\
1
\end{array}
\right)
.
\end{eqnarray}
For the second line to hold everywhere on the plane and for all times, we need
\begin{equation}
\mathbf{k}\cdot\mathbf{x}|_{\mathbf{x}=(x,y,z=0)}
=\mathbf{p}\cdot\mathbf{x}|_{\mathbf{x}=(x,y,z=0)}.
\end{equation}
This gives
\begin{equation}
\mathbf{k}_{||}=\mathbf{p}_{||},
\end{equation}
and determines two of the three components of $\mathbf{k}$.
The remaining component is determined by energy conservation, which imposes 
\begin{equation}
|\mathbf{k}|=\omega=\sqrt{\mg^2+|\mathbf{p}|^{2}}.
\end{equation}
Explicitly we have,
\begin{equation}
\mathbf{k}=\sqrt{\mg^2+|\mathbf{p}_{\perp}|^{2}}\mathbf{n}+\mathbf{p}_{||},
\end{equation}
where $\mathbf{p}_{\perp}$ is the component of $\mathbf{p}$ perpendicular to the surface and $\mathbf{n}$ is the unit vector normal to the surface.

This means that for non-relativistic momenta, $|\mathbf{p}|\ll \mg $ of the incoming dark matter particles the produced electromagnetic waves are emitted at a small angle $\psi \simeq |\mathbf{p}_{||}|/\mg$ with respect to the normal of the surface.

When the wavelength is much smaller than the size of the surface in question we can use the ray approximation. This is what we will do in the following.

\subsection*{Spherical cap}

\begin{figure}[h]
\begin{center}
\vspace{-2cm}
\includegraphics[width=15cm]{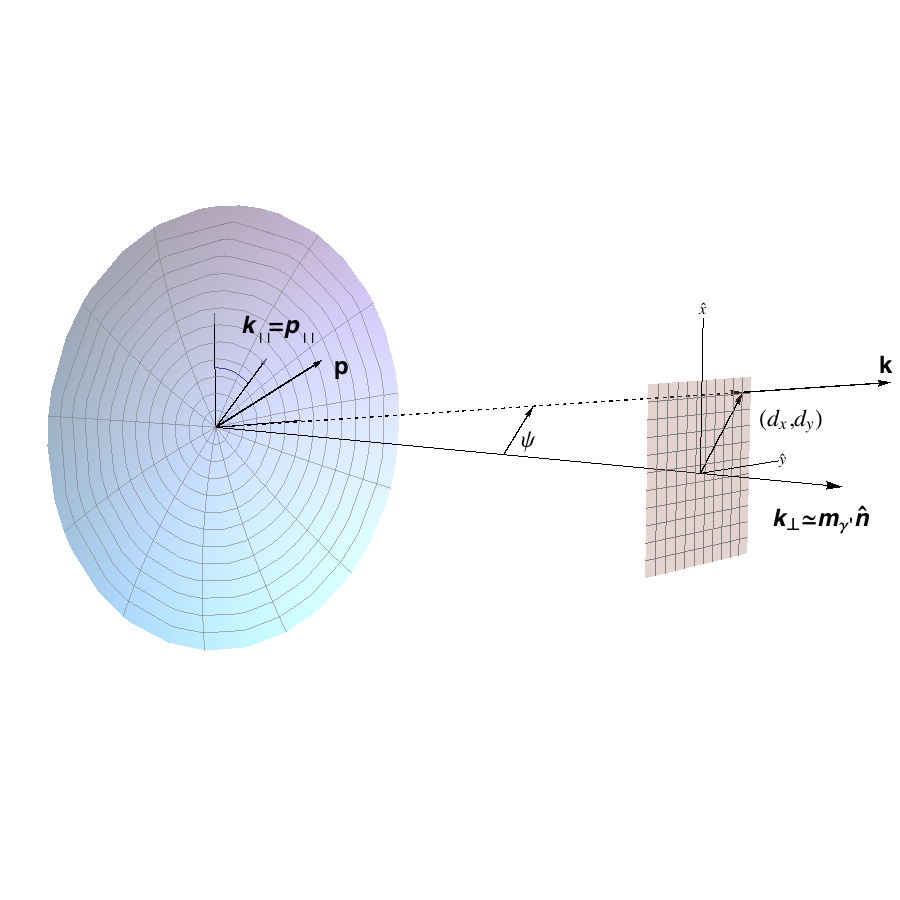}
\vspace{-4cm}
\caption{Schematic of our directional DM detection. Reflecting spherical dish with a parallel planar detector with position sensitivity located in the center of the sphere. 
In response to a HP DM with mass $m$ and small momentum $\mathbf{p}=(\mathbf{p}_{||},\mathbf{p}_{\perp})$ (parallel and perpendicular to the surface at a given point) the mirror emits photons with momentum $\mathbf{k}\simeq (\mathbf{p}_{||},m \mathbf{ n})$ where $\mathbf n$ is the normal to the surface. The emission angle with respect to $\mathbf{n}$ is $\psi\simeq |\mathbf{p}_{||}|/m$.  
Photons emitted from the center of the sphere hit the detector at coordinates $(d_x,d_y)\simeq R \mathbf{p}_{||}/m $ with respect to the detector center ($R$ is the radius of the sphere). }
\label{setup}
\end{center}
\end{figure}

Let us now consider a dish antenna shaped like a spherical cap and a detector with sensitive area centered in the centre of the sphere and perpendicular to the axis of revolution of the cap, see Fig.~\ref{setup}.
For vanishing $\mathbf{p}_{||}$ all rays are concentrated in the centre of the sphere, as already discussed in~\cite{Horns:2012jf}. Let us now investigate what happens for the case of a general non-vanishing $\mathbf{p_{||}}$ (but still small $|\mathbf{p}|\ll \mg$).

Let us first consider a ray emitted from the centre of the dish as shown in Fig.~\ref{setup}. At leading order in 
$\mathbf{p}/\mg$
this ray will hit the detector at
\begin{equation}
d_{x}=\frac{p_{x}}{\mg}R,\quad d_{y}=\frac{p_{y}}{\mg} R.
\end{equation}
In other words the position of the signal in the detector plane is directly proportional to the momentum of the incoming dark matter particle. 
This suggests that the signal strength allows us to directly reconstruct the momentum distribution in the $x$ and $y$
directions.

\begin{figure}[h]
\begin{center}
\vspace{-2cm}
\includegraphics[width=15cm]{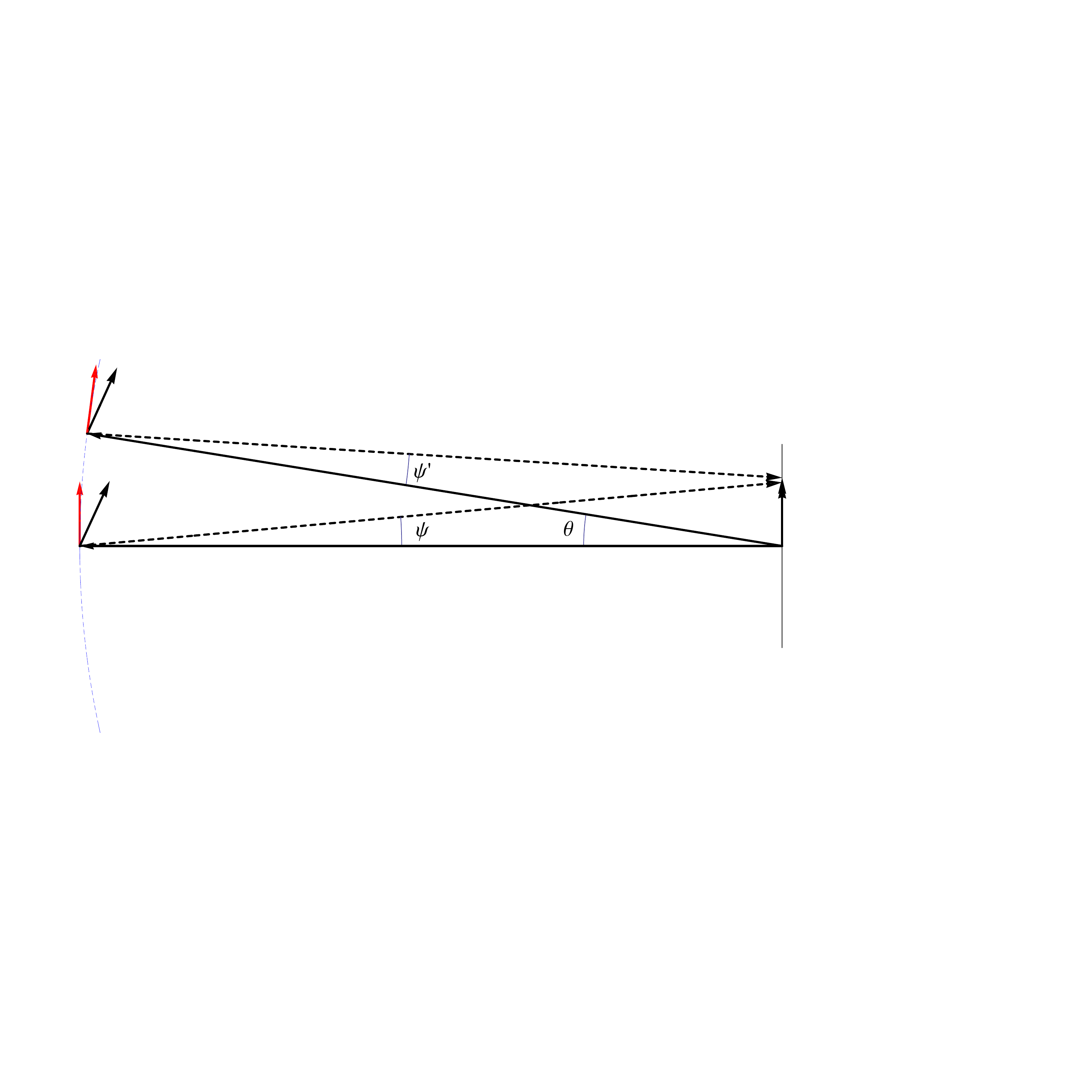}
\vspace{-4cm}
\caption{In the ray approximation, photons emitted from different regions of the dish deviate from the perpendicular by an angle $\psi\simeq\mathbf{p}_{||}/m$ where $\mathbf{p}_{||}$ (red vector) is the component of the DM momentum $\mathbf{p}$ (black vector). 
Different points in the sphere feel the same $\mathbf{p}$ but have different $\mathbf{p}_{||}$. 
In a region of polar coordinate $\theta$ (assumed to be small) the normal to the surface rotates by an angle $\sim \theta$ and thus the parallel component $\mathbf{p}_{||}$ is increased by a factor $\sim p_{\perp}\theta$. The angle of photon emission is slightly different for different $\theta$, $\psi'(\theta)-\psi(0)\simeq\theta p_\perp/m$. 
Thus, light rays arrive at the detector at positions which differ by $\Delta d\simeq R \Delta \psi \simeq R \theta p_\perp/m.$
}
\label{setup3}
\end{center}
\end{figure}

To get an appreciable signal it is crucial that all rays emitted from different points on the sphere are concentrated in the same place of the detector. A short geometrical calculation shows that the rays originating at an arbitrary point of the spherical cap 
--denoted by spherical coordinates $(\theta,\phi)$, polar and azimuth-- reach the detector at the point
\begin{equation}
d^{\prime}_{x}\simeq R\left(\frac{p_{x}}{\mg}-\frac{p_{z}}{\mg}\tan(\theta)\cos(\phi)\right),\quad 
d^{\prime}_{y}\simeq R\left(\frac{p_{y}}{\mg}-\frac{p_{z}}{\mg}\tan(\theta)\sin(\phi)\right).
\end{equation}
to leading order in $\mathbf{p}/\mg$. 
In Fig.~\ref{setup3} we schematically explain the origin of the dispersion in a 2-D example.  

Thus, a ray originating at polar coordinate $\theta$ ends up at a distance $\Delta d=R \tan(\theta) p_z/m$ from the one originating in the center ($\theta=0$). 
In order to get rid of this aberration we should consider dishes with an aperture radius $r$ much smaller than the radius of curvature, i.e. we demand 
\begin{equation}
\theta_{\rm max}\approx r/R\ll 1. 
\end{equation} 
The resolution in momenta is thus at the level 
\begin{equation}
\label{resolution}
\frac{\Delta p}{p_{\rm typical}}\sim\theta_{\rm max}\approx \frac{r}{R}, 
\end{equation}
where $p_{\rm typical}\sim10^{-3}$ is the typical momentum of the dark matter particles.

It is clear that for fixed area of the dish ($A_{\rm dish}
\simeq \pi r^2$) the best resolution can be obtained for a fairly large curvature radius $R$.

Using now the probability distribution $f(\mathbf{p})$ in momentum space we can immediately determine the signal strength
at the position $(d_{x},d_{y})$ of the detector (neglecting the small smearing of order $\theta_{\rm max}$) with the power per area given by
\begin{equation}
\frac{d{P}(d_{x},d_{y})}{d A}=P_{0}\frac{\mg^{2}}{R^2}\int \frac{dp_{z}}{(2\pi)^{3}} f\left(\frac{d_{x}}{R}\mg,\frac{d_{y}}{R}\mg,p_{z}\right),
\end{equation}
with 
\begin{equation}
\label{totalpower}
P_{0}\approx \langle\cos^2\alpha\rangle_{\rm dish} \chi^2\rho_{\rm CDM, halo} A_{\rm dish},
\end{equation}
the total power emitted by the dish~\cite{Horns:2012jf}.
Here $\alpha$ is the angle between the HP polarization and the surface of the dish. The average takes into account the different hidden photon orientations with respect to the dish and is given by,
\begin{equation}
\langle\cos^2\alpha\rangle_{\rm dish}=\bigg\{
\begin{array}{cl}
\cos^2(\alpha_0) & {\rm scenario\,\, (i)}\\
\sqrt{\frac{2}{3}}& {\rm scenario\,\, (ii)}
\end{array},
\end{equation}
where $\alpha_0$ is the angle between the hidden photon field and the dish surface in scenario (i) (we assumed $r\ll R$)
and the average in Eq.~\eqref{totalpower} is taken over the surface of the dish antenna. 

\subsection{Further limitations on the directional resolution}
We have already seen above that the resolution is limited by the fact that for large dishes the outer parts of the dish focus a the electromagnetic wave caused by a moving hidden photon into a slightly different spot than the central parts of the dish.
The resolution limit is given by Eq.~\eqref{resolution}. This limitation is purely geometrical and can already be seen in the ray approximation we have employed so far.
We have seen that it can be reduced by using a large radius of curvature.

\subsubsection{Diffraction}
A second limitation arises from diffraction.
At fairly large wavelengths which are relevant for the lower frequency/mass part of the range for which our setup can be employed, diffraction will provide a severe limit to the ability for directional detection.
Diffraction limits the angular resolution to,
\begin{equation}
\Delta \psi \gtrsim \frac{\lambda}{2 r}.
\end{equation}
For our directional detection this causes an uncertainty on the DM velocity 
\begin{equation}
\frac{\Delta \mathbf{p}}{\mg}\sim \frac{\lambda}{r} \sim10^{-6}\left(\frac{\rm eV}{\mg}\right)\left(\frac{\rm m}{r}\right).
\end{equation}
For the relative resolution this entails,
\begin{equation}
\label{diffraction}
\frac{\Delta \mathbf{p}}{p_{\rm typical}}\sim 10^{-3}\(\frac{10^{-3}}{v_{\rm typ}}\)\left(\frac{\rm eV}{\mg}\right)\left(\frac{\rm m}{r}\right),
\end{equation}
since the typical velocity in the galaxy is expected to be $v_{\rm typ}\sim 10^{-3}$.

As we can see diffraction is not a problem in the optical and near infrared even if we want to achieve a relative precision in 
the \% range.
For radio frequencies, i.e. masses below $100\,{\rm meV}$, however this is a serious limitation that requires very large dishes and, because of Eq.~\eqref{resolution} even larger radius of curvature. 

\subsubsection{Earth's motion}
Let us define the velocity distribution in the galactic rest frame, where is expected to have regular properties, 
\begin{equation}
f_{\rm Galactic\,\, frame}(\mathbf{v'}), 
\end{equation}
where primed velocities are defined in the Galactic rest frame. 
The Lab frame on Earth where we perform our dish experiment is moving together with the rest of the solar system at a velocity $\mathbf{v}'_{\rm Earth}$ of magnitude $\sim 0.7\times 10^{-3}$ (220 km/s) around the galactic center. 
But the Earth spins around itself with a period of one day around the South-North direction ($l,b\sim123\degree,27\degree.4$ in galactic coordinates). Thus, in a Lab frame co-rotating with the Earth the velocity distribution is
\begin{equation}
f_{\rm Earth\,\, frame}(\mathbf{v})\equiv f_{\rm Galactic\,\, frame}(\mathbf{v}')=f_{\rm Galactic\,\, frame}(R(t)\mathbf{v}+\mathbf{v}'_{\rm Earth}).
\end{equation}
where $R(t)$ is the time-dependent rotation matrix transforming velocities in the Lab frame to the galactic frame. 

This introduces two effects. First there is an offset in the detector due the relative motion of the Earth to the dark matter wind. Assuming the DM velocity distribution to be isotropic in the Galactic frame, on Earth we see it biased towards the direction $-\mathbf{v}_{\rm Earth}$, which means that the signal in our detector will be centered around the point $(d_x,d_y)=R({v}_{\rm Earth,x},{v}_{\rm Earth,y})$. Since this direction is known, one can easily correct for this bias. 
Second, this direction is time-dependent $\mathbf{v}_{\rm Earth}=R^{-1}(t)\mathbf{v}'_{\rm Earth}$. 
It precedes around the North-pole with an angle $\beta\sim 42\degree$ and a period of a day. 
As an example, if the axis of revolution of our dish coincides with the South-North pole axis, the signal focus will describe a circle of radius $0.7\times10^{-3}\times R\cos(\beta)$ around the center of the detector. For a more general orientation, the focus describes ellipsoids around a different displaced point. 

Again, this can be corrected in at least two ways: a) one can mount the detector on some movable device that shifts the detector to track the offset or b) one can split the data taking period in time-frames where the displacement of the focus is smaller than the required angular accuracy and combine them in the final analysis correcting for the time-dependent offset. 
For an angular accuracy of $\Delta \psi$ the frames should be shorter than
\begin{equation}
\label{timeframes}
t_{\rm min}\sim \frac{\rm day}{|\mathbf{v}_{\rm Earth}|} {\Delta \psi} \sim 
 20\, {\rm min} \frac{\Delta \psi}{10^{-5}\rm rad} \ . 
\end{equation} 

\subsection{Energy spectrum}
So far we have focussed mainly on determining the momentum distribution of the dark matter particles.
More precisely, we have determined the velocity distribution of the dark matter particles, $\mathbf{v}=\mathbf{p}/\mg $.   
Indeed one can easily check that \emph{to determine the velocity distribution} as described above \emph{it is not necessary to know the mass $\mg$}.
The mass $\mg\approx \omega$ can be inferred from spectroscopy of the signal. 

Spectroscopy may also allow us to go beyond just determining the mass and determine the spectrum of the kinetic energy. This works exactly in the same manner as in resonant cavity experiments~\cite{Duffy:2005ab}. 

Let us nevertheless note a couple of points. 
First of all, high resolution spectroscopy is relatively simple for radio frequencies. There typical measurement devices will already do spectroscopy in order to detect only a narrow peak and to suppress thermal and other backgrounds that grow with the bandwidth. Obtaining a high resolution spectrum of a peak then just requires a little bit of extra measurement time.
This is nicely complementary, as we have seen in the previous subsection directional detection is more difficult for low frequencies.
In this case one can then at least obtain the distribution of the velocity squared or the modulus of the velocity.

Second, for higher frequency (optical or infrared photons) high precision spectroscopy is probably slightly more involved,
as one will need to add an additional spectrometer to the setup. Moreover, these measurements will typically yield
only very few photons, providing an additional challenge for spectroscopy.
 
\subsection{Determining the direction of the HP vector}\label{HP-direction}
Hidden photons, being vector particles have an intrinsic directionality.
So in principle we can have different distributions for the three vectorial components.

There are two ways to obtain information about this directionality. The first is that according to Eq.~\eqref{totalpower} the total signal varies with the relative orientation of the direction of the HP field and the antenna. After all only HP field components parallel to the antenna plane can be converted into ordinary photons. 
However, this is only sensitive to an average of the HP directions, i.e. the net directionality.

A second way provides more detailed information on the directionality: we can measure the polarization of the produced photons.
Employing a polarization filter in front of the detector one can get the signal strength for each direction separately. Combining this with a detector that has spatial resolution, as discussed above, we can actually measure the velocity spectrum for each component of the vector field.

Note, that these considerations do not apply to axions which, being scalar particles, have no intrinsic directionality.
There the polarization is entirely determined by the external magnetic field used for the conversion of axions into photons.

\section{A simple example setup}\label{example}
To demonstrate the viability and to gain insight into the challenges of directional detection let us study an example setup.
Let us consider a simple toy distribution, 
\begin{equation}
\label{distribution}
f_{\rm Gal}(\mathbf{v}')={\mathcal N} \Theta(v^{2}_{\rm max}-|\mathbf{v}'|^2),
\end{equation}
with $v_{\rm max}= 10^{-3}= 300\,{\rm km}/{\rm s}$. 
${\mathcal N}$ is a normalization constant such that the density is $0.3$ GeV/cm$^3$ 

Let us then consider the following experimental parameters for our setup,
\begin{equation}
{\rm Mirror:}\,\,r=1\,{\rm m}, \quad R=10\,{\rm m}.
\end{equation}
We point the axis of the dish ($z$) such that the velocity of Earth lies perpendicular to it, along the $x$-axis 
\begin{equation}
v_{{\rm Earth},x}=0.7\times 10^{-3} =220 \,{\rm km/s},\quad v_{y,\rm Earth}=v_{z,\rm Earth}=0.
\end{equation}
The area in which we expect the signal is a circle of radius $10^{-3}R=10 $ mm 
displaced by a distance $v_{{\rm Earth},x}R=7$ mm from the center along the $x$-direction.   
We then take as detector a CCD of total size 
\begin{equation}
{\rm Detector}:\,\, 40\, {\rm mm}\times 40\,{\rm mm},
\end{equation}

\begin{figure}[t]
    \begin{center}
    \vspace*{0.5cm}
         \includegraphics[width=8cm]{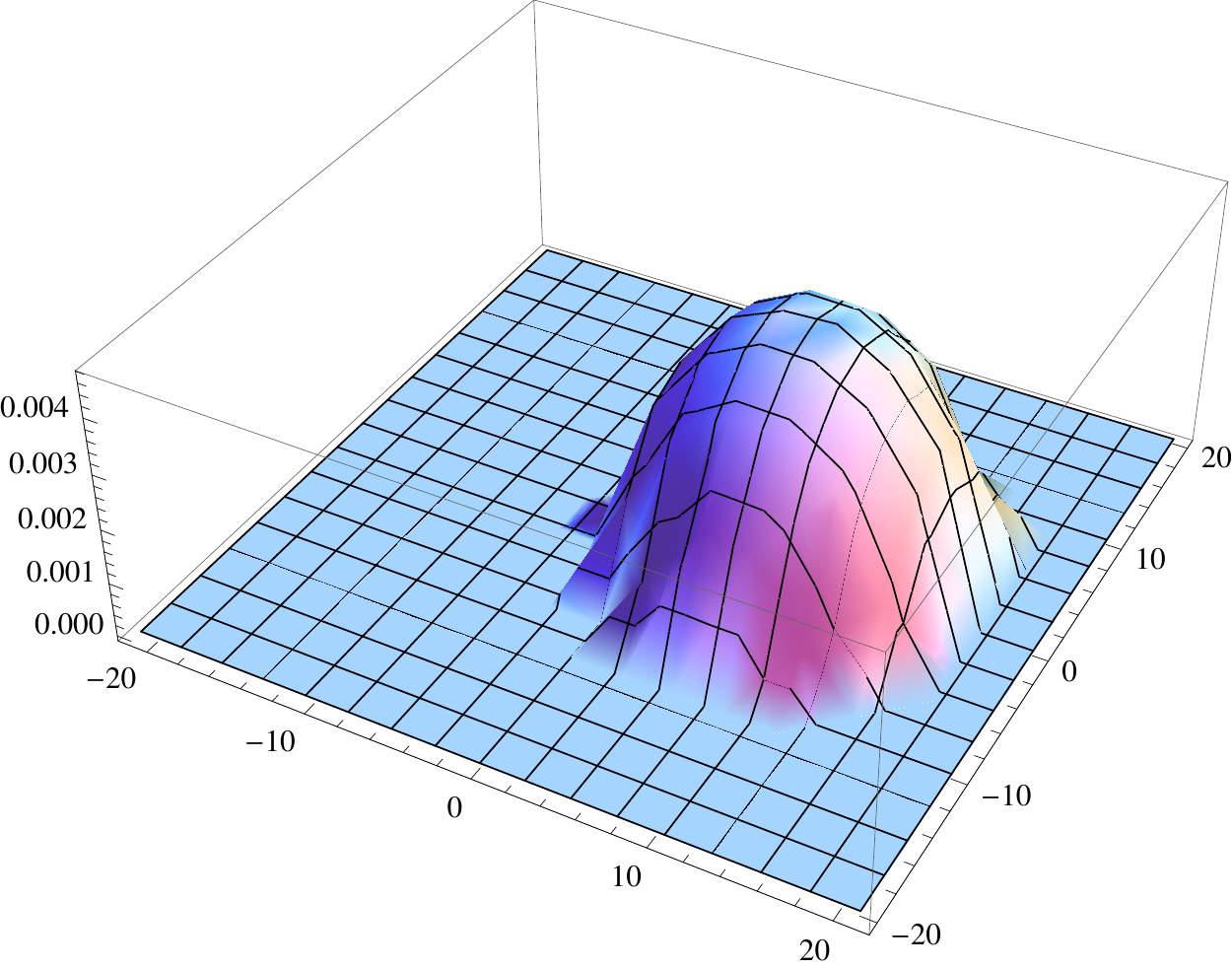}
        \caption{Distribution of the dark matter signal in the detector plane (location in mm) for a dark matter signal described by Eq.~\eqref{distribution} and a geometry as described in the text. The offset from the center is caused by the velocity of Earth through the dark matter relative to the direction in which the dish antenna is pointing.}
\label{demonstration}
    \end{center}
\end{figure}

From Eqs.\eqref{diffraction} and~\eqref{resolution} it is clear that the blurring caused by diffraction is negligible ($\Delta d = R\Delta\psi\sim R \lambda/ r\sim 10\mu$m) compared to the one caused by the non-ideal mapping of the outer regions of the dish, $\Delta d \lesssim R \theta_{\rm max} v_{\rm typical}\sim r v_{\rm typical} \sim 0.7$ mm. 
The resolution of the detector needs not being smaller than 1mm so we take  
\begin{equation}
{\rm Pixels}:\,\, 1\, {\rm mm}\times 1\,{\rm mm}, {160\, \rm pixels}
\end{equation}
In Fig.~\ref{demonstration} we show the resulting Power distribution in the detector's plane for this exemplary set-up.  
We note that the distribution we obtain is averaged over the velocity in the unresolved  $z$-direction (that's why it also does not look like a step function).

Overall it is clear that we have quite good directional sensitivity. In particular we note that we can clearly see the offset caused by the Earth's movement through the dark matter wind.

\bigskip
For practical considerations the most important point may be signal strength. 
Let us now consider dark matter with
\begin{equation}
\mg=1\,{\rm eV}\quad  \chi= 3\times 10^{-12},
\end{equation}
corresponding the strongest interactions compatible with current bounds~\cite{An:2013yua,Redondo:2013lna}.  
 
We expect a total signal photon rate 
\begin{equation}
\dot{N}^{\rm total}_{\gamma}\approx 2.5\times \langle \cos^2\alpha\rangle \, \rm Hz , 
\end{equation}
These photons are rougly spread out over a circle with a radius of 10~mm. Therefore in the signal region we expect,
\begin{equation}
\frac{\dot{N}_{\gamma}}{\rm area}\sim 8\times \langle \cos^2\alpha\rangle\frac{\rm mHz}{mm^2}.
\end{equation}

If order to avoid smearing of the signal further than the 1 mm due to precession of $\mathbf{v}_{\rm Earth}$,  time-frames should be shorter than $\sim 3$ hours (From using $\Delta\psi <0.1 $mm in Eq.~\ref{timeframes}). 

After these 3 hours we have $\sim 80$ signal photons in each bin. Measurements can be repeated roughly within these time-scales to reduce backgrounds. All in all, the measurement is quite challenging but after all one would gain an enormous amount of detailed information.
\
\section{Effects on discovery experiments}
Experiments aiming at a first detection of HP dark matter do not need to achieve directional sensitivity. However, taking the effects of the dark matter velocity into account is nevertheless important.

From the discussion in the previous sections it is clear that not all photons are concentrated in a single spot but rather there is a finite distribution in the detector plane. In other words it is crucial to have a sufficiently sized detector in order to collect a sufficient amount of signal.

Assuming that we have a velocity distribution with a width of
\begin{equation}
\Delta v\sim 300\,{\rm km}/{\rm s}\sim 10^{-3}
\end{equation}
we expect that the extend of the velocity distribution in the detector plane is of the order of
\begin{equation}
\Delta d\sim \Delta v R\sim 1\,{\rm mm}\left(\frac{R}{{\rm m}}\right).
\end{equation}

\begin{figure}[t]
    \begin{center}
    \vspace*{0.5cm}
         \includegraphics[width=8cm]{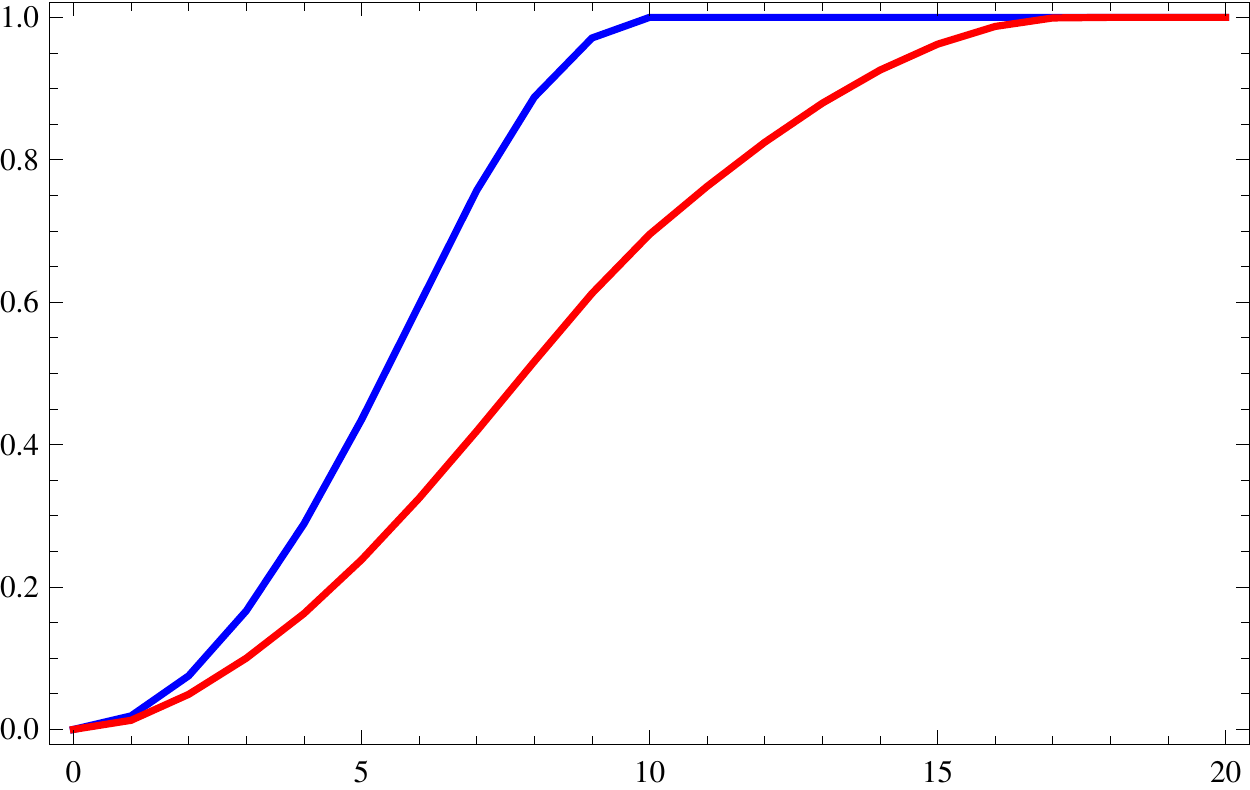}
        \caption{Fraction of the total signal power captured by the detector as a function of the size (in mm) of the detector for the example setup of Sect.~\ref{example}.}
\label{fraction}
    \end{center}
\end{figure}

This is a quite sizeable effect. For our example setup from the previous section we show the fraction of signal hitting a detector depending on the detector size in Fig.~\ref{fraction} as the black curve. 
Indeed, this plot actually assumes that we always point the detector into the direction with which Earth is moving through the halo, thereby centering the signal on the center of the detector.

If we do not take care to point our detector into the dark matter wind, we need an even larger detector due to the offset of the dark matter signal from the center as can be seen in Fig.~\ref{demonstration}. The blue curve in Fig.~\ref{fraction} shows the fraction of signal hitting the detector for the relative orientation without correcting for the Earth's velocity (for our example situation of Sect.~\ref{example}).

Taking the Earth's velocity into account may be non-trivial. In order to collect enough signal we may have to integrate for fairly long times possible of the order of a day or more. On these time-scales the rotation of the Earth around its axis is non-negligible. Therefore, one has to either track the Earth's rotation or the signal is spread out over a larger area. Again the signal fraction hitting the detector for our example setup is shown as the red curve in Fig.~\ref{fraction}.

The required size of the detector is also very relevant for the fundamental backgrounds. 
The maximal dish size $r_{\rm dish}$ is the curvature radius $R$. In other words
\begin{equation}
\frac{A_{\rm detector}}{A_{\rm dish}}\gtrsim \Delta v^2,
\end{equation}
in order to collect an appreciable part of the signal.
This shows that when increasing the size of the antenna we also need to increase the size of the detector.
Accordingly our signal to background ratio only improves with increasing dish size when the noise in the detector grows slower than the area.

\section{Conclusions}
Dish antennas can be a powerful tool to search for WISPy dark matter coupled to photons such as hidden photons and axion like particles. In this note we have demonstrated that this technique can not only be used as a discovery experiment but once discovery is made, directional detection, giving information on the full vectorial velocity distribution can be done in a straightforward manner, opening a path to dark astronomy.

Importantly, taking into account that dark matter has a non-vanishing velocity is also crucial in designing and optimizing discovery experiments, even those which do not aim for directional detection, by requiring a sufficient size of the detector. 

\section*{Acknowledgements}

J.R. acknowledges support by the Alexander von Humboldt Foundation and 
partial support by the European Union through the Initial Training Network ``Invisibles''.

\begin{appendix}
\section*{Appendix: The 0-component of the vector field}
The equations of motion for the massive vector field automatically enforce the Lorentz gauge condition. For the massless field this can be done by a suitable gauge choice. Let us therefore use,
\begin{equation}
\partial_{\mu}A^{\mu}=0,\quad \partial_{\mu}X^{\mu}=0.
\end{equation}
For a wave solution,
\begin{equation}
\sim \exp(-i(\omega t-\mathbf{p}\cdot\mathbf{x})),
\end{equation}
we then have,
\begin{equation}
A^{0}=\frac{\mathbf{p}\cdot\mathbf{A}}{\omega},\quad X^{0}=\frac{\mathbf{p}\cdot\mathbf{X}}{\omega}.
\end{equation}

We now have a massive solution,
\begin{equation}
\left(\begin{array}{c}
A^{\mu}\\
X^{\mu}
\end{array}
\right)= X^{\mu}_{0}\,(\mathbf{p})\exp(-i(\omega t-\mathbf{p}\cdot\mathbf{x}))
\left(
\begin{array}{c}
-\chi\\
1
\end{array}\right)
\end{equation}
and a massless one
\begin{equation}
\left(\begin{array}{c}
A^{\mu}\\
X^{\mu}
\end{array}
\right)= A^{\mu}_{0}\,(\mathbf{p})\exp(-i(\omega t-\mathbf{p}\cdot\mathbf{x}))
\left(
\begin{array}{c}
1\\
\chi
\end{array}\right),
\end{equation}
where we have, in some abuse of notation, written the coefficients according to the dominant components.
The DM solution is the massive one.

We now have to implement the boundary conditions.
As in the main text we consider a perfectly conducting plane at $z=0$. 

Let us consider a DM particle that has no velocity in the x-direction, i.e. $p_{x}=0$ (this can always be done without loss of generality, by rotating around the z-axis). 
We can now decompose,
\begin{equation}
X^{\mu}_{0}=
X_{x}\left(
\begin{array}{c}
0\\
1\\
0\\
0
\end{array}\right)
+X_{y}\left(
\begin{array}{c}
-\frac{p_{y}}{\omega}\\
0\\
1\\
0
\end{array}\right)
+X_{z}\left(
\begin{array}{c}
-\frac{p_{z}}{\omega}\\
0\\
0\\
1
\end{array}
\right)
\end{equation}
such that each individual bit fulfils the Lorentz gauge condition.

The first bit is a vector component parallel to the plane but perpendicular to the velocity of the DM particle.
This part has no 0-component and the treatment of Sect.~\ref{directsect} is exact.

For the remaining two parts we now have to implement the boundary condition 
\begin{equation}
E_{y}=0
\end{equation}
by adding a suitable multiple of the massless field to cancel the electric field in the plane.
In analogy to the massive mode
we now have two vectors perpendicular to the x-direction which fulfil the Lorentz condition.
\begin{equation}
\left(
\begin{array}{c}
-\frac{k_{y}}{\omega}\\
0\\
1\\
0
\end{array}\right)
\quad{\rm and} \quad\left(
\begin{array}{c}
-\frac{k_{z}}{\omega}\\
0\\
0\\
1
\end{array}
\right).
\end{equation}
However, one can quickly check (using that for the massless mode we have $\omega^2=k^{2}_{y}+k^{2}_{z}$)
that both lead to the same physical field configuration,
\begin{equation}
\left(
\begin{array}{c}
E_{x}\\
E_{y}\\
E_{z}
\end{array}
\right)
\sim
\left(
\begin{array}{c}
0\\
k_{z}\\
-k_{y}
\end{array}
\right)
\end{equation}
which is the proper transversal mode with $\mathbf{E}\perp \mathbf{k}$.
The difference between the two is a gauge mode. Indeed if the velocity is perpendicular to the plane, $k_{y}=0$,
the first mode again has vanishing 0-component and the second is a pure gauge mode with vanishing physical fields.
Therefore we can simply pick the first one and use it to fulfil the boundary condition,
\begin{equation}
A^{\mu}_{0}=A_{y}\left(
\begin{array}{c}
-\frac{k_{y}}{\omega}\\
0\\
1\\
0
\end{array}\right).
\end{equation}
This field gives electric fields,
\begin{equation}
\label{field}
\left(
\begin{array}{c}
E_{x}\\
E_{y}\\
E_{z}
\end{array}
\right)
=A_{y}\left(i\frac{k_{z}}{\omega}\right)
\left(
\begin{array}{c}
0\\
k_{z}\\
-k_{y}
\end{array}
\right)
\end{equation}
For the boundary condition we then have (using that to match the phases in the plane we need $k_{y}=p_{y}$),
\begin{equation}
0=E_{y}=i\frac{k^{2}_{z}}{\omega}A_{y}-i\chi\left[\left(\omega-\frac{p^{2}_{y}}{\omega}\right)X_{y}-\frac{p_{y}p_{z}}{\omega}X_{z}\right].
\end{equation}
From this we can then determine
\begin{equation}
A_{y}=\chi \left[X_{y}-\frac{p_{y}p_{z}}{\omega^2-p^{2}_{y}}X_{z}\right]=\chi X_{y}+{\mathcal{O}}(v^{2}).
\end{equation}
Using Eq.~\eqref{field} one can then quickly check that the intensity of the outgoing wave is also only modified at order $v^2$.

\end{appendix}

\end{document}